\documentclass[aps,prl,twocolumn,preprintnumbers,amsmath,amssymb]{revtex4-1}

\usepackage{graphicx}
\usepackage{bm}
\usepackage{upgreek}

\begin{document}

\title{Quantization of the superconducting energy gap in an intense microwave field}

\author{A. A. Boris}
\author{V. M. Krasnov}
\email{Vladimir.Krasnov@fysik.su.se}

\affiliation{ Department of Physics, Stockholm University,
AlbaNova University Center, SE-10691 Stockholm, Sweden }

\date{\today}

\begin{abstract}
We study photon assisted tunneling in Nb/AlO$_x$/Nb Josephson
junctions. A quantitative calibration of the microwave field in
the junction allowed direct verification of the quantum efficiency
of microwave photon detection by the junctions. We observe that
voltages of photon assisted tunneling steps vary both with the
microwave power and the tunneling current. However, this variation
is not monotonous, but staircase-like. The phenomenon is caused by
mutual locking of positive 
and negative 
step series. A similar locking is observed with Shapiro steps. As
a result the superconducting gap assumes quantized values equal to
multiples of the quarter of the photon energy. The quantization is
a manifestation of nonequilibrium tuning (suppression or
enhancement) of superconductivity by the microwave field.

\end{abstract}


\maketitle


Superconducting Josephson junctions are used as sensitive
detectors of microwave (MW) and terahertz signals
\cite{Tucker_1985,Likharev_1986,Hu_1990,Goltsman_1996,Koshelets_2000,Divin_2000,Basset_2012,Malnou_2014}.
Application of MW radiation leads to appearance of Shapiro and
photon assisted tunneling (PAT) steps in the current-voltage
($I$-$V$) characteristics of a junction at $eV = \pm n hf/2$ and
$eV = \pm 2\Delta \pm n hf$, respectively. Here $f$ is the MW
frequency and $\Delta$ is the superconducting energy gap. Shapiro
and PAT steps originate from Cooper pair and quasiparticle (QP)
currents, correspondingly (for details see e.g.
Ref.~\cite{Barone}). A response of junctions to weak MW signals is
well understood \cite{Tucker_1985,Likharev_1986,Hu_1990}. For
example, the differential conductivity due to PAT is described by
the Tien-Gordon theory:
\begin{equation}\label{PAT}
\frac{dI}{dV}(V)=\sum^{\infty}_{n=-\infty}{J_n^2\left(\frac{eV_{MW}}{hf}\right)\frac{dI_0}{dV}\left(V+\frac{nhf}{e}\right)},
\end{equation}
where $J_n$ are Bessel functions of integer order $n$, $V_{MW}$ is
the MW voltage amplitude in the junction and $I_0(V)$ is a
dc-current without MW. Positive/negative $n$ terms in this
expression describe contributions from PAT with
absorbtion/emission of $n$ photons.

However, such a textbook description works only for weak MW
signals that do not affect superconducting properties of junction
electrodes. Strong electromagnetic fields may disturb a thermal
equilibrium state. This may either suppress
\cite{Winkler_1987,Gershenzon_1990} or enhance
\cite{Wyatt_1966,Clarke_1977,Horstman_1981,Beck_2013}
superconducting properties, such as $\Delta$ and the critical
current $I_c$. Current flow through the junctions also leads to
disruption of the equilibrium. Analysis of such nonequilibrium
effects at low temperatures is complicated by essentially
nonlinear response of the junction to perturbations
\cite{Krasnov_2009}. Detailed understanding of nonequilibrium
effects in Josephson junctions is lacking and is important both
for optimization of operation of superconducting quantum devices
\cite{Martinis_2013,Pekola_2013,Kubatkin_2013,Klapwijk_2014,Devoret_2014,Delsing_2014}
and for fundamental studies of the pairing mechanism in
unconventional superconductors \cite{Krasnov_2013}. To investigate
the influence of nonequilibrium effects in intense MW fields $V_{MV}>2\Delta/e$ on
detection characteristics of Josephson junctions is the main
objective of our work.

In this work we study photon assisted tunneling steps in
Nb/AlOx/Nb Josephson junctions. Using an absolute calibration of
the MW field {\it inside} the junction we demonstrate quantum
efficiency of microwave photon detection by our junctions. We
observe that contrary to expectations PAT voltages are not
constant, but vary both with the MW power $P_{MW}$ and the current
through the junction. However, this variation is not monotonous,
but staircase-like. The phenomenon is caused by mutual locking of
positive $eV=2\Delta -nhf$ and negative $eV=-2\Delta +mhf$ PAT
series at large MW amplitudes $eV_{MW} \gtrsim \Delta$. 
A similar locking is observed between Shapiro and PAT steps. The
locking is a manifestation of nonequilibrium adjustment of the
superconducting gap. As a result, $\Delta$ assumes quantized
values equal to multiples of the quarter of the photon energy.

We study micron-size Nb/AlO$_x$/Nb junctions made by a standard
technology. Several junctions with different sizes on the same
chip are studied. Detailed characterization of our junctions can
be found in Ref.~\cite{MR}. Measurements are performed in a
pulsed-tube cooled optical cryostat. A single loop frequency
synthesizer with a 8x multiplier is used as a MW source. The MW
power is supplied to the junction quasioptically using a system of
high-density polyethylene lenses. The MW power $P_{MW}$ is tuned
using two polarizers. The polarizer at the source site is rotated
by a step motor. The polarizer at the sample site is aligned with
the source and is fixed. The applied MW power is measured
using an opto-acoustical detector (Golay cell).

\begin{figure*}[t]
    \centering
    \includegraphics[width=0.95\textwidth]{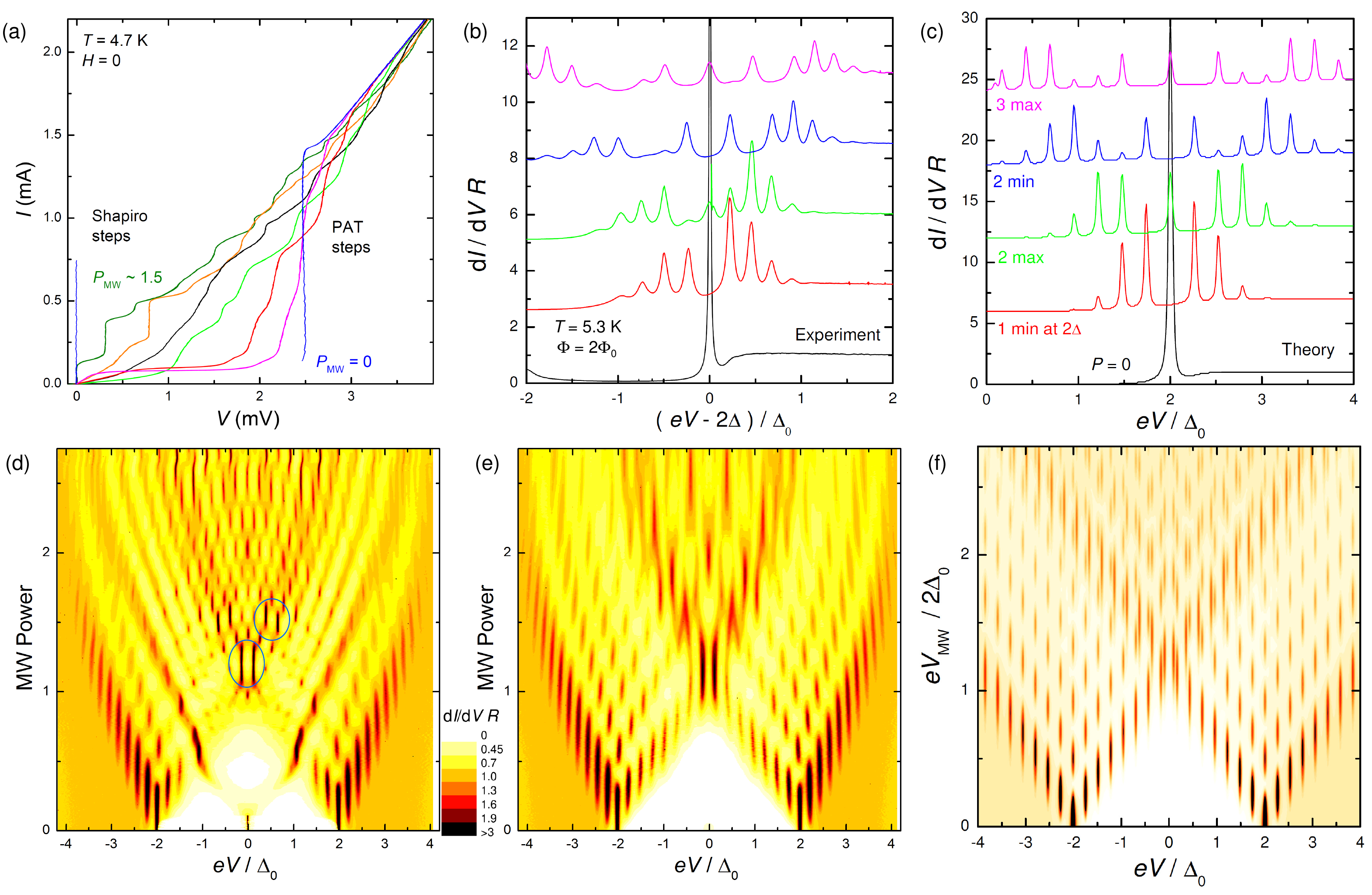}
    \caption{(Color online). (a) Current-voltage characteristics at $H=0$ and $T=4.7$ K for different
    $P_{MW}=0$, 0.15, 0.275, 0.463, 0.825, 1.25, 1.5 at $f=76.6$
    GHz. (b) Experimental and (c) theoretical $dI/dV(V)$ curves with PAT steps for
    several $P_{MW}$ corresponding to maxima and minima of the sum-gap step $eV=2\Delta$.
    Curves are offset vertically for clarity.
    (d) and (e) Color maps of differential conductances for the same junction at $T=5.3$ K for (d) $H=0$ and
    (e) $\Phi=2\Phi_0$. Both Shapiro and PAT
    steps are seen in (d), but only PAT steps are present in (e). (f)
    Theoretical color map of $dI/dV$ for PAT as a function of bias voltage $V$ and the MW voltage inside the junction
    $V_{MW}$. Note that positive and negative step series overlap at $eV_{MW}/2\Delta_0
    \simeq 1$. }
    \label{fig:fig1}
\end{figure*}

Figure 1 (a) shows $I$-$V$ characteristics of a junction at zero
magnetic field for several $P_{MW}$ at $f\simeq 76.6$ GHz. With
increasing $P_{MW}$ PAT steps at $eV = \pm 2\Delta \pm n hf$ and
Shapiro steps at $eV = \pm n hf/2$ appear.  Step amplitudes oscillate with $P_{MW}$.
The supercurrent can be suppressed by applying magnetic field parallel to the junction
plane, corresponding to integer number of flux quanta $\Phi_0$ in
the junction. In this case only PAT steps are present.
Fig.~\ref{fig:fig1} (b) represents $dI/dV(V)$ curves, normalized
by the tunnel resistance $R\simeq 3.6~ \Omega$, at $\Phi =
2\Phi_0$ for a slightly smaller junction. Curves are presented for
MW powers corresponding to maxima and minima of the sum-gap step
$eV=2\Delta$.
In Fig.~\ref{fig:fig1}(c) we show corresponding theoretical
calculations according to Eq. (\ref{PAT}) \cite{note} (curves for
different $P_{MW}$ in Figs. \ref{fig:fig1} (b) and (c) are offset
vertically for clarity).

Figs.~1(d) and (e) represent color maps of $dI/dV~R$ as a function
of bias voltage and MW power.
Measurements are performed at $T=5.3$ K for (d) $H=0$ and (e) at
$\Phi=2\Phi_0$. At $H=0$ both Shapiro and PAT steps are present.
At $\Phi=2\Phi_0$ solely the PAT step structure is seen. Fig.~1(f)
shows a corresponding theoretical calculation for PAT, obtained
from Eq.~(\ref{PAT}), as a function of the MW voltage. It is seen
that with increasing $V_{MW}$ two PAT step series start to expand
in a V-shape manner from the sum-gap peaks at positive $eV=2\Delta
\pm nhf$ and negative $eV=-2\Delta \pm mhf$ voltages. The steps
modulate quasi-periodically with increasing $V_{MW}$. At $eV_{MW}
= 2\Delta$ photon absorption steps from the two series overlap. We
use this point for absolute calibration of the MW amplitude {\it
inside} the junction. In what follows we normalize $P_{MW}$ to
this value.

\begin{figure*}[t]
    \centering
    \includegraphics[width=0.95\textwidth]{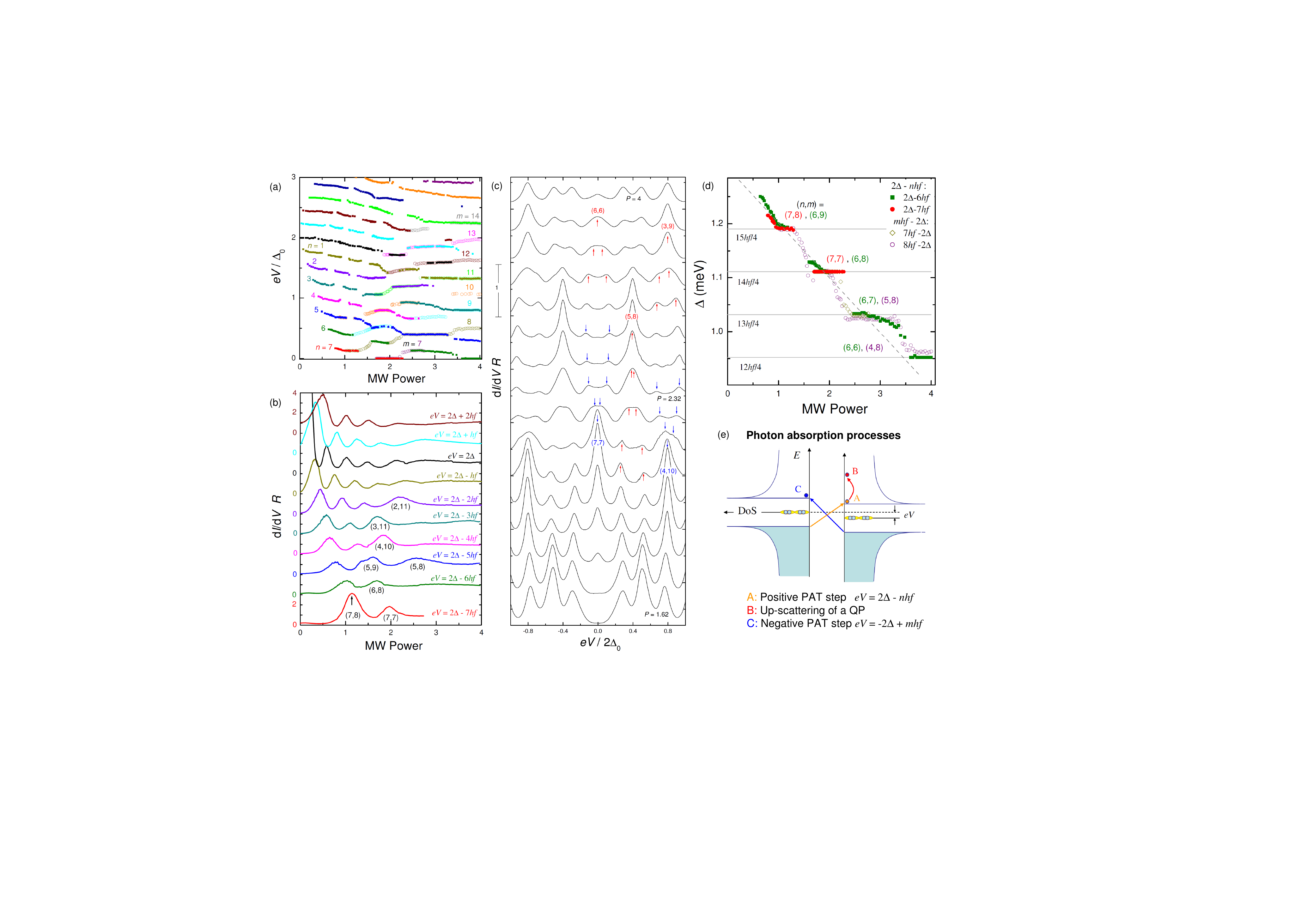}
    \caption{(Color online). (a) PAT step voltages versus MW power from Fig.~\ref{fig:fig1}(c).
    Solid and open symbols represent positive and negative PAT step series with orders $n$ and $m$, respectively.
    Note locking of those steps upon overlap.
    (b) Normalized step amplitude as a function of MW power.
    Numbers $(n,m)$ indicate formation of high amplitude steps due to locking
    of the corresponding positive $n$ and negative $m$ steps.
    (c) $dI/dV(V)$ curves at $P_{MW}$ from 1.62 to 4. Up/down
    arrows indicate merging/splitting of PAT steps, leading to
    appearance/dissapearance of high amplitude steps.
    Curves in (b) and (c) are shifted vertically for clarity.
    (d) The superconducting energy for several PAT steps. Quantization of the gap at $P_{MW}>1$ is seen.
    (e) Diagrams of various nonequilibrium processes, associated with photon absorption. }
    \label{fig:fig2}
\end{figure*}

Comparison of theory with experiment reveals a qualitative
agreement, but also certain differences. First, in theory PAT
steps oscillate quasi-periodically with $V_{MW}$, while in
experiment with $P_{MW}$ \cite{Supplem}. This may indicate that we
have reached the quantum efficiency of photon detection
\cite{Tucker_1985}, in which case the number of tunneled electron
is limited by the number of incoming photons. The quantum-limited
current at the $n$-th photon absorption step is $I_n = (e/n)
dN/dt$, where $dN/dt$ is the rate of incoming photons, connected
to the absorbed MW power $P_{a}=hf dN/dt$. Thus, such current is
proportional to the {\it absorbed} MW power rather than $V_{MW}$,
$I_n=P_{a}e/nhf$. Usually, however, it is difficult to identify
exactly how much MW power is absorbed by the junction. Therefore,
to verify this idea we propose to use the absolute calibration of
the MW field {\it inside} the junction at the overlap point
$V_{MW}(P_{MW}=1)\simeq 2\Delta/e$, which yields
$P_{a}=V_{MW}^2/4R=P_{MW}(\Delta^2/e^2R)$. Here an additional
factor $1/2$ is due to rectifying action of the junction due to
which at positive dc-bias the MW current is flowing only during
positive half-periods of oscillations and $P_{MW}$ is the
normalized MW power to the overlap point value. Thus,
quantum-limited $n$-photon current is
$I_n=P_{MW}(\Delta/nhf)(\Delta/eR)$. For the case of Fig.
\ref{fig:fig1} (a), $\Delta\simeq 2.5$ mV, $\Delta/hf\simeq 3.9$
and the sum-gap kink amplitude $2\Delta/eR\simeq 1.4$ mA, we get
$I_n\simeq (2.7/n) P_{MW}$ (mA). Magenta line in Fig.
\ref{fig:fig1} (a), $P_{MW}\simeq 0.15$, corresponds to the first
maximum of the single photon step, at which contribution from
multi-photon tunneling is small. It is seen that the amplitude of
the single photon absorption step, $eV=2\Delta-hf$, is very close
to the calculated $I_1\simeq 0.4$ mA, confirming the quantum
efficiency of detection of MW photons by our junctions.

Another difference between experiment and theory, which is in
focus of this work, occurs at high MW powers $P_{MW}>1$ when
positive and negative PAT steps overlap. In theory,
Fig.~\ref{fig:fig1} (f), this does not lead to anything special.
But in experiment, Fig.~\ref{fig:fig1} (e), it clearly causes a
complete reconstruction of the step pattern. Instead of having two
independent series of steps, the nearby steps from positive and
negative sides lock together forming fewer steps with
significantly stronger amplitudes. A similar locking effect is
seen in Fig.~\ref{fig:fig1}(d) between Shapiro and PAT steps,
which leads to appearance of super-steps, marked by ovals. Such
super-steps do not occur in the absence of PAT \cite{Malnou_2014}.

Figure~\ref{fig:fig2} (a) shows PAT step voltages from
Fig.~\ref{fig:fig1} (e), normalized by the gap $\Delta_0$ at
$P_{MW}=0$. It is seen that contrary to a textbook case step
voltages are neither constant nor equidistant. Since PAT voltages
are strictly connected to the gap, $eV_n = \pm 2\Delta \pm nhf$,
variation of step voltages implies variation of $\Delta$.
Generally, step voltages decrease with increasing $P_{MW}$. This
indicates suppression of the gap by the MW field. We emphasize
that the base temperature was actively controlled during the
experiment and was constant in the whole range of $P_{MW}$ with
the accuracy better than 0.1 K. Thus, suppression of $\Delta$ is
due to nonequilibrium over-population of the electronic system
\cite{Krasnov_2009}, rather than overheating of the system.
Suppression of $\Delta$ leads to a decrease of positive $eV_n =
2\Delta - nhf$ (solid symbols) and increase of negative $eV_m =
-2\Delta + mhf$ (open symbols) photon absorption steps. Thus,
positive and negative steps gradually approach each other and
overlap at $P_{MW}>1$. As seen from Fig. \ref{fig:fig1} (e), the
overlap leads to a complete reconstruction of the PAT pattern at
$P_{MW}>1$: nearby positive and negative steps lock-in together to
form high-amplitude steps with constant (power-independent)
voltages.

In Fig. \ref{fig:fig2} (b) we show amplitudes of PAT steps from 
Fig. \ref{fig:fig2} (a) of different $n$ as a
function of the MW power (symbols and lines with the same color in Figs.  \ref{fig:fig2} (a) and (b) 
represent steps of the same order). Periodic modulation is clearly seen (see
also the Supplementary \cite{Supplem}). We mark steps formed by
locking of $n$-th order positive and $m$-th order negative photon
absorption steps by doublets $(n,m)$. They have significantly
higher amplitudes than foregoing maxima with the same $n$, or than
corresponding maxima of the proceeding order $n-1$. For example
the third maximum for $n=3$, marked (3,11), is due to locking
between $n=3$ and $m=11$ steps. It has a larger amplitude than
both the second $n=3$ and the third $n=2$ maxima. Fig.
\ref{fig:fig2} (c) shows evolution of $dI/dV(V)$ curves in the
range of $P_{MW}$ from 1.62 to 4. Up/down arrows indicate
merging/splitting of PAT steps, leading to
appearance/dissapearance of high amplitude steps $(n,m)$.

A condition for locking of positive and negative photon absorption
steps is $ 2\Delta - nhf = mhf - 2\Delta$. Therefore, upon such
locking the gap assumes quantized values,
\begin{equation}\label{Delta}
    \Delta = (n+m)hf/4.
\end{equation}

Fig.~\ref{fig:fig2} (d) shows the energy gap vs.  MW
power for several PAT steps. The gap is obtained explicitly as $\Delta=(eV_n+nhf)/2$ for positive and $\Delta=(mhf-eV_m)/2$ negative step series.
A staircase-like variation of
$\Delta(P_{MW})$ with platos at quarters of the photon
energy is clearly seen. The quantization becomes progressively
stronger with increasing the MW power. Ovals in Fig.~\ref{fig:fig1} (d) indicate that a similar locking takes
place between Shapiro and PAT steps. This occurs when $nhf/2 =
2\Delta - mhf$ leading to the same type of quantization,
Eq.~(\ref{Delta}). Locking between Shapiro and PAT steps suggests that the locking
involves collective ac-Josephson currents both for pairs and quasiparticles, such as the illusive QP interference current (the cosine term) \cite{Barone}. 

Finally we want to discuss the mechanism of the observed gap
quantization. Fig.~\ref{fig:fig2} (e) represents diagrams of three
possible photon absorption processes. The process A corresponds to
positive PAT steps. It leads to breaking of a Cooper pair and
appearance of nonequilibrium electron and hole-like QPs, which
suppresses $\Delta$. The process B corresponds to up-scattering of
a quasiparticle. According the the self-consistency equation,
$\Delta$ is most sensitive to QPs just at the bottom of the gap
(see e.g. Ref.~\cite{Krasnov_2009}). Therefore the process B leads
to enhancement of the gap. The process C corresponds to negative
PAT steps. On one hand, it leads to enhanced nonequilibrium QP
population and thus decreases $\Delta$, on the other hand the
current is flowing against the voltage, leading to negative
dissipation power and cooling of the junction, which enhances
$\Delta$. Thus there are several nonequilibrium processes in the
MW field that may either increase or decrease $\Delta$. Such a
tunability is required for locking of PAT steps. The dashed line
in Fig.~\ref{fig:fig2} (d) indicates that the gap 
in average is decreasing approximately linearly with increasing
$P_{MW}$. However at the steps it deviates both up and down from
this line showing that both enhancement and suppression of the gap
indeed takes place.

To conclude, quantitative calibration of the microwave field in
Josephson junction allowed direct verification of the quantum
efficiency of microwave photon detection by the junctions. Our
main result is observation of a quantization of the
superconducting gap in intense microwave fields. The phenomenon is
caused by mutual locking of photon assisted tunneling steps due to
nonequilibrium enhancement and suppression of superconductivity by
the microwave field.

\begin{acknowledgments}
The work was supported by the Swedish Foundation for International
Cooperation in Research and Higher Education Grant No. IG2013-5453
and the Swedish Research Council Grant No. 621-2014-4314.
\end{acknowledgments}

\end {document}